\documentclass[reprint,amsmath,amssymb,aps,superscriptaddress,prl]{revtex4-2}

\usepackage{graphicx}
\usepackage{xcolor}
\usepackage{dcolumn}
\usepackage{bm}
\usepackage{booktabs}
\usepackage{sidecap}
\sidecaptionvpos{figure}{t}
\usepackage[pdftex,bookmarks,colorlinks,breaklinks]{hyperref} 
\hypersetup{linkcolor=red,citecolor=blue,filecolor=dullmagenta,urlcolor=blue} 
\hypersetup{pdftoolbar=true,pdfpagemode=UseNone,pdfstartview=FitH, colorlinks=true,extension=}

\begin{document}

\title[Synchronization driven reciprocity breaking]{Synchronization driven reciprocity breaking}
\author{Alexander K. Stoychev}
\email[]{astoychev@ethz.ch}
\affiliation{CAPS Laboratory, Department of Mechanical and Process Engineering, ETH Zürich, 8092 Zürich, Switzerland}

\author{Ulrich Kuhl}
\affiliation{CAPS Laboratory, Department of Mechanical and Process Engineering, ETH Zürich, 8092 Zürich, Switzerland}
\affiliation{Université Côte d’Azur, CNRS, Institut de Physique de Nice (INPHYNI), 06200, Nice, France}

\author{Nicolas Noiray}
\email[]{noirayn@ethz.ch}
\affiliation{CAPS Laboratory, Department of Mechanical and Process Engineering, ETH Zürich, 8092 Zürich, Switzerland}

\date{\today}
\begin{abstract}
This study introduces a novel method to break wave transmission reciprocity by leveraging the synchronization of self-oscillators. An experimental demonstration with aeroacoustic cavities is presented. They behave as weakly nonlinear limit cycles when driven by a constant airflow, leading to self-oscillations which can couple to the surrounding waveguides via two ports. Incident waves from one port trigger anti-phase synchronization, causing destructive interference and low transmission, while waves from the opposite port induce in-phase synchronization, resulting in high transmission. This directional dependence effectively breaks reciprocity, where the operational bandwidth is defined by the synchronization region (Arnold tongue), and can be broader than resonance based methods. Experimental results show robust nonreciprocal behavior w.r.t. parameter changes. Moreover, a modified temporal coupled mode theory is proposed, explaining the system’s nonlinear dynamics and scattering properties in a quantitative manner. This synchronization-based approach offers a new avenue for directional wave control, complementing traditional reciprocity breaking techniques, and offering an intrinsic loss-compensation emanating from the self-oscillation of meta-atoms.
\end{abstract}
\maketitle

\emph{Introduction -- }
Lorentz reciprocity in passive linear wave media dictates that source and receiver are interchangeable under the assumptions of (microscopic) time-reversal symmetry and time-invariance, implying that an unidirectional transmission (e.g. diodes, insulators, circulators, etc.) requires bias or nontrivial material properties. In particular, strategies to break reciprocity have been characterized \cite{Nassar2020} into: (i) momentum bias (breaking time-reversal) \cite{Kittel1958,Fleury2014,Nash2015,Kariyado2015,Wiederhold2019}, (ii) spatio-temporal modulations (breaking time-invariance) \cite{Hu2006,Wang2013,Zanjani2014,Nassar2017,Wang2018,Nassar2017a,Quan2019,Liu2019,Cheng2022,Wang2023,Nassar2018,Chen2019,Brandenbourger2019,Zhang2021,Zhang2023}, and (iii) structural asymmetry with nonlinearity (breaking linearity) \cite{Liang2009,Liang2010,Boechler2011,Coulais2017,Lu2020}. To achieve nonreciprocity in kinetic medium, a momentum bias could be introduced via circulating fluids \cite{Fleury2014}, spinning masses \cite{Nash2015,Wiederhold2019}, Coriolis
forces \cite{Kariyado2015}, magnetic fields \cite{Kittel1958}, etc. Conversely, spatio-temporal modulations can manifest in directional Bragg reflection \cite{Hu2006,Wang2013,Zanjani2014,Nassar2017,Wang2018}, Willis coupling \cite{Nassar2017a,Quan2019,Liu2019,Cheng2022,Wang2023} or topologically protected edge states \cite{Nassar2018,Chen2019,Zhang2021,Zhang2023}. Finally, reciprocity breaking nonlinear effects could range from a static response of an asymmetric structure \cite{Coulais2017}, to generation of harmonics and band-pass filtering \cite{Liang2009,Liang2010}, bifurcation driven transitions \cite{Boechler2011}, self-demodulation effects \cite{Devaux2015}, or topological solitons~\cite{Veenstra2024}.
\begin{figure}
    \includegraphics[width=\linewidth]{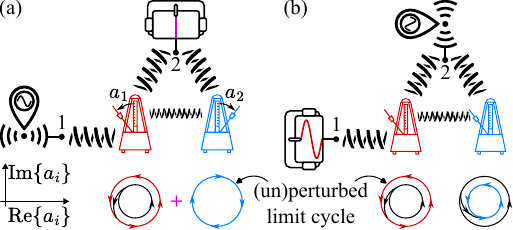}
    \caption{Principle of the synchronization driven reciprocity breaking. (a) Incidence from port 1 couples to one of the limit cycles (red) and due to the non-linear interaction of the self-oscillators (SO), both of them contribute to the transmission. (b) Incidence from port 2 forces both SOs simultaneously, while only red SO determines the transmission. The direct coupling between the ports is weak by design, consequently, the (forced) synchronization state governs the scattering properties of the system. The lower part indicates phase space behavior of the limit cycles upon incidence from the outside.
    }\label{fig:Concept}
    \vspace{-5mm}
\end{figure}
\vspace{-1mm}\newline\vspace{0mm}
We propose a fundamentally different approach, which relies on weakly nonlinear limit cycles and direction dependent synchronization rather than bias, band topology, or harmonics. Importantly, the combination of an isolated mode (separated from other resonances) and a saturable-gain mechanism, lead to single degree-of-freedom self-oscillators which, despite nonlinear effects, remain quasi-linear in spectrum (minimal harmonic content). Two such coupled oscillators (one per waveguide port) then interact with the incident waves, where crucially, the scattering is determined by the synchronization state [Fig.~\ref{fig:Concept}]. Specifically, an incident wave from port 1 primarily interacts with one limit cycle, while both settle into an anti-phase mode with their emissions canceling, which leads to low transmission [Fig.~\ref{fig:Concept}(a)]. Conversely, an incident wave from port 2 drives both oscillators simultaneously, allowing them to lock in phase and radiate constructively, yielding high transmission [Fig.~\ref{fig:Concept}(b)]. The resulting synchronization pattern governs the reciprocity breaking, where the bandwidth is set by the width of the synchronization region, while the weak nonlinearity filters out small perturbations. Note that the two SOs don't need to be precisely adjusted but it is sufficient that the driving frequency-amplitude pair is in the range of the joint synchronization region. 

\begin{figure}
    \includegraphics[width=\linewidth]{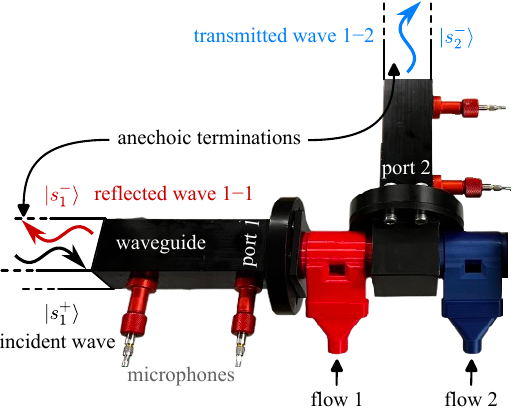}
    \caption{Aeroacoustical implementation of the reciprocity breaking self-excited meta-atom assembly. Only an incidence from port 1 is depicted here, where an incidence from port 2 follows analogously [cf. Fig.~\ref{fig:Concept}]. The apertures on top of the cavities add irreversible losses in the passive case (no limit cycle), and saturable gain in the self-excited case (limit cycle). The cavities (whistles) are coupled to the wave guides via small perforations on the sides, where the blue one is coupled only to port 2.}
    \label{fig:ExperimentalSetup}
    \vspace{-4mm}
\end{figure}

\emph{Results -- }
In the following we focus on an aeroacoustic implementation [Fig.~\ref{fig:ExperimentalSetup}] of the synchronization based approach discussed above. Specifically, a limit cycle is implemented through a self-oscillating cavity mode, pumped by the self-induced conversion of an externally supplied, steady flow's kinetic energy into an acoustic one \cite{LimitCycleTransistor}. Note that in this case the underlying mode derives from the constant pressure solution of a rigid cavity, and therefore, is (i) incommensurate with any higher order eigenfrequencies, and (ii) resulting in a wavelength large compared to the dimensions of the meta-atom. Nevertheless, those properties are not essential to the proposed method and we could consider non-compact scatterers by working with higher order modes of the whistle, taking advantage of $\pi$ phase shifts at different port locations. Crucially, however, the response is only weakly nonlinear and nonreciprocity is achieved at the frequency of the incident wave. Consequently, the whistles [Fig.~\ref{fig:ExperimentalSetup}] behave as a single degree of freedom self-oscillators suitable for averaging \cite{Krylov1950}, allowing us to accurately model them with a first order, nonlinear ODE for the complex amplitude~$a$~\cite{LimitCycleTransistor}. 

The limit cycles are coupled to the surrounding waveguides (and each other) via small perforations in the cavity side walls, which allows them to synchronize with the incident wave, and with each other. Since the resonant coupling (no limit cycle) is weak enough to render the individual oscillators mostly reflective \cite{LimitCycleTransistor}, the scattering properties are determined by the synchronization state of the system. The latter can be tuned by the distance between the whistles, which corresponds to the coupling phase of the SOs. Moreover, varying the position of the source [Figs.~\ref{fig:Concept}\&\ref{fig:ExperimentalSetup}] could result in (i) the self-sustained anti-phase mode being excited by manipulating the synchronization phase of the first oscillator via the first port, leading to phase canceling and low transmission, or to (ii) the in-phase mode being instigated by forcing both oscillators to synchronize with a wave incident from the second port, leading to high transmission. Analogous to the scattering by a single element \cite{LimitCycleTransistor}, the latter can exhibit (i) lossy, (ii) unitary, or (iii) superradiant behavior.

Note that the synchronized states we are interested in are constrained to specific regions in parameter space, concretely, the ones where the incident amplitude is large enough compared to the frequency detuning [the Arnold tongue]. Inside the aforementioned areas the spectrum of the resulting signal is (nearly) monochromatic, where outside thereof we find at least two significant components; (i) at the self-oscillation in addition to (ii) the omnipresent incident frequency, with (iii) even richer response close to synchronization (cf. \cite{LimitCycleTransistor} App.~B).

\emph{Experimental Measurements -- }
\begin{SCfigure*}
    \includegraphics[width=360pt]{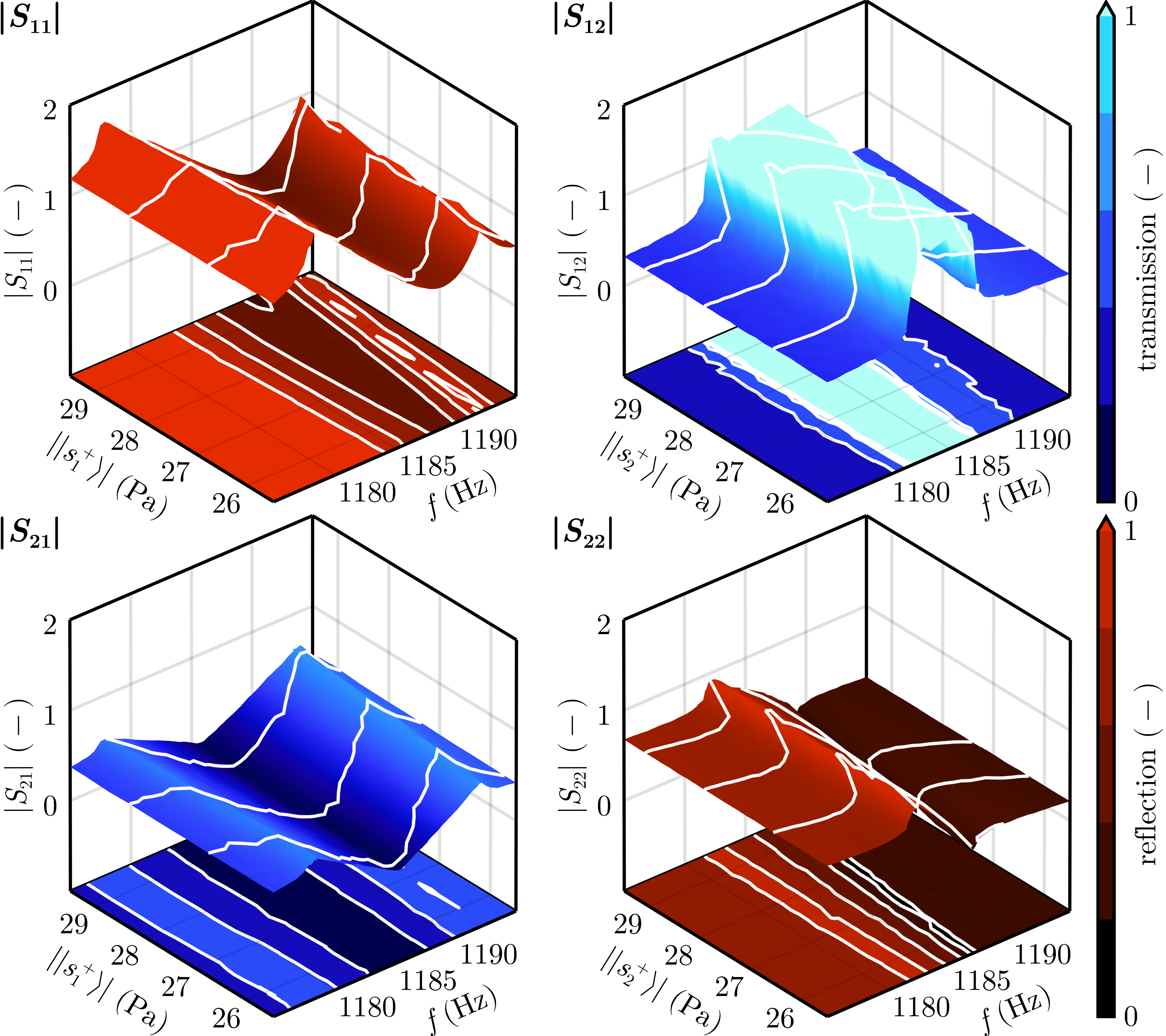}
    \caption{Experimental measurements of the transmissions [off-diagonal; $S_{12}$ \& $S_{21}$] and reflections [diagonal; $S_{11}$ \& $S_{22}$] as function of incident frequency, $f$, and amplitude, $\left\vert\left\vert s^\mathrm{+}\right\rangle\right\vert$, for an incidence from port 1 [left; $S_{11}$ \& $S_{21}$] and port 2 [right; $S_{12}$ \& $S_{22}$], at flow rates of $\dot{Q}_1=11\,\mathrm{l/min}$ and $\dot{Q}_2=11.25\,\mathrm{l/min}$ for the red and blue whistle (see Fig.~\ref{fig:ExperimentalSetup}), respectively. The scattering data shows only slight dependence on the incident wave amplitude, signifying a weak nonlinearity and leading to persistent reciprocity breaking across the considered range. Conversely, the frequency, $f$, has strong effect on the reflection and a less prominent, yet significant, effect on the transmission. Nevertheless, the bandwidth of the process is governed by the synchronization region, which can be influenced, e.g., by adjusting the coupling coefficients. Note that a loudspeaker input signal varying solely the frequency $f$ results in some incident amplitude variation due to reflections from the terminations [white lines in 3D plot].}
    \label{fig:ScatteringExperiment}
\end{SCfigure*}
Fig.~\ref{fig:ScatteringExperiment} illustrates the concept, where the four sub-panels depict the corresponding entries of the scattering matrix, each as a function of the incident wave's frequency and amplitude.

In particular, comparison between the transmission from port 1 to port 2 [$S_{21}$] and the reverse [$S_{12}$], i.e., the off-diagonal elements, shows the expected unidirectional channeling close to the self-oscillation frequency [$f_0\approx1187\,\mathrm{Hz}$]. Although the scattering properties of the system are intrinsically affected by the incident amplitude due to nonlinear effects \cite{LimitCycleTransistor}, this dependence is relatively week, leading to a comparatively robust to perturbations reciprocity breaking mechanism. Specifically, over the considered range we observe slightly superradiant transmission in one direction [$S_{12}>1$] and (approximate) phase canceling in the other [$S_{21}\approx0$], where the bandwidth of the former is inherited from the Arnold tongue \cite{LimitCycleTransistor} and the one of the latter is determined by the phase relations of the synchronization state.


\emph{Theoretical model -- }
To model the experimentally observed behavior we adopt a temporal coupled mode theory (TCMT) approach \cite{Fan2003}, although, with several necessary modifications, i.e.,
\begin{align}
  &~\begin{aligned}
      \frac{\mathrm{d}a_1}{\mathrm{d}\tau} = \left(j\frac{1-\Omega^2}{2\Omega}\right.&+\left.\frac{\mu}{2}\frac{1-\vert a_1\vert^2}{1+\vert a_1\vert^2}\right)a_1 \\
    &+\varkappa a_2 + \left\langle\kappa_1\big\vert s^{\mathrm{+}}\right\rangle\,,
  \end{aligned} \label{eq:LimitCyclePathway1} \\
  &~\begin{aligned}
      \frac{\mathrm{d}a_2}{\mathrm{d}\tau} = \left(j\frac{1-\Omega^2}{2\Omega}\right.&+\left.\frac{\mu}{2}\frac{1-\vert a_2\vert^2}{1+\vert a_2\vert^2}\right)a_2 \\
      &+\varkappa a_1 + \left\langle\kappa_2\big\vert s^{\mathrm{+}}\right\rangle\,,
  \end{aligned} \label{eq:LimitCyclePathway2} \\
  &\left\vert s^{\mathrm{-}}\right\rangle = \mathbf{D}\left\vert s^{\mathrm{+}}\right\rangle + \left\vert \tilde\kappa_1\right\rangle a_1 + \left\vert \tilde\kappa_2\right\rangle a_2\,, \label{eq:DirectPathway}
\end{align}
where $\mathrm{d}/\mathrm{d}\tau$ represents derivative w.r.t.\ time, normalized by the self-oscillation frequency [$f_0\in\mathbb{R}$], while $j$ denotes the imaginary unit, and $\Omega=f/f_0\in\mathbb{R}$ is the ratio between intrinsic [$1/f_0$] and forcing time scale [$1/f$]. Furthermore, $a_{1\{2\}}\in\mathbb{C}$ characterizes the complex amplitude of the corresponding limit cycle, with $\vert a_{1\{2\}}\vert$ and $\arg\{a_{1\{2\}}\}$ describing the slow time scale magnitude and phase behavior, respectively. Moreover, $\mu\in\mathbb{R}$ incorporates the radiation and viscous losses, $\varkappa=|\kappa_1||\tilde\kappa_2|\exp{(j\phi)}=|\kappa_2||\tilde\kappa_1|\exp{(j\phi)}\in\mathbb{C}$ relates to the inter-limit-cycle interactions, and $\left\langle\kappa_1\right\vert=\kappa_1\left\langle1\right\vert+\kappa_1\left\langle2\right\vert$, $\left\langle\kappa_2\right\vert=\kappa_2\left\langle2\right\vert$, with $\kappa_{1\{2\}}\in\mathbb{R}$, stand for the direct coupling matrices. Additionally, $\left\vert s^{\mathrm{+}}\right\rangle = s\left\vert1\{2\}\right\rangle$ indicates the incident wave, with $s=\left\vert\left\vert s_p^{\mathrm{+}}\right\rangle\right\vert/\left\vert p_p\right\vert\in\mathbb{R}$ being the amplitude of the physical wave, $\left\vert\left\vert s_p^{\mathrm{+}}\right\rangle\right\vert$, normalized by the unperturbed limit cycle pressure $\left\vert p_p\right\vert$. Finally, the complex valued $\mathbf{D}\approx\mathbf{1}$ signifies the scattering matrix of the direct process and $\left\langle\tilde\kappa_1\right\vert=\tilde\kappa_1\left\langle1\right\vert+\tilde\kappa_1\left\langle2\right\vert$, $\left\langle\tilde\kappa_2\right\vert=\tilde\kappa_2\left\langle2\right\vert$, with $\tilde\kappa_{1\{2\}}\in\mathbb{R}$, denote the conjugate coupling matrices.
\begin{SCfigure*}
    \includegraphics[width=360pt]{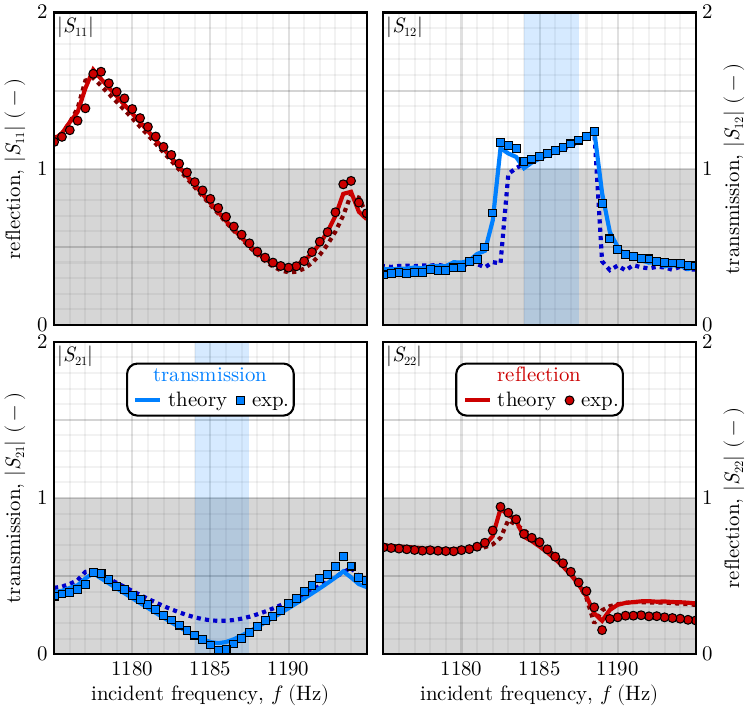}
    \caption{Scattering coefficients as a function of frequency, $f$, for an incident amplitude of $\left\vert\left\vert s_1^{\mathrm{+}}\right\rangle\right\vert\approx\left\vert\left\vert s_2^{\mathrm{+}}\right\rangle\right\vert\approx29\,\mathrm{Pa}$ and flow rates of $\dot{Q}_1=11\,\mathrm{l/min}$ and $\dot{Q}_2=11.25\,\mathrm{l/min}$, respectively. Left panels refer to incidence from port 1 [$S_{11}$ \& $S_{21}$], while right panels correspond to waves incident from port 2 [$S_{21}$ \& $S_{22}$; cf. Fig~\ref{fig:ExperimentalSetup}]. Blue colors represent transmission, with square markers depicting measurement points and solid (dotted) lines - theoretical calculations. Conversely, red colors denote reflection, with circular markers being measurements and solid (dotted) lines - solutions of Eqs.~\eqref{eq:LimitCyclePathway1}--\eqref{eq:DirectPathway}. Dotted lines correspond to $\kappa_2=\kappa_1=\kappa$ and solid ones to $\kappa_2=1.12\kappa$, where the latter is shown to be responsible for the transmission jump and bandwidth widening in $S_{12}$. Note that despite keeping the external excitation within each datasets constant, the actual incident amplitude varies slightly with frequency due to back-reflections from the terminations [cf. 3D lines in Fig.~\ref{fig:ScatteringExperiment}]. Therefore, the results (theoretical and experimental) depict a flat projection of curved trajectories through parameter space.}
    \label{fig:ScatteringTheory}
\end{SCfigure*}
\begin{table*}
    \vspace{-3mm}
    \centering
    \caption{Reduced model parameters.}
    \label{tab:ReducedModelParameters}
    \setlength{\tabcolsep}{8pt}
    \begin{tabular}{ccccccccccc}
        \toprule
        $\!\left\vert p_p\right\vert~(\mathrm{Pa})$ & $f_0~(\mathrm{Hz})$ & $D_{11}~(\mathrm{-})$ & $D_{12}~(\mathrm{-})$ & $D_{21}~(\mathrm{-})$ & $D_{22}~(\mathrm{-})$ & $\mu~(\mathrm{-})$ & $\tilde\kappa~(\mathrm{-})$ & $\kappa~(\mathrm{-})$ & $\arg\varkappa~(\mathrm{rad})\!$ \\
        \midrule
        $350$ & $1186.9$ & $0.9\cdot\mathrm{e}^{2.6j}$ & $0.05\cdot\mathrm{e}^{-0.08j}$ & $0.06\cdot\mathrm{e}^{0.92j}$ & $0.8\cdot\mathrm{e}^{2.5j}$ & $0.125$ & $0.072$ & $0.208$ & $3.075$ \\
        \bottomrule
    \end{tabular}
    \vspace{-3mm}
\end{table*}
\vspace{-1.2mm}\newline\vspace{0mm}
Analogous to the single limit cycle description \cite{LimitCycleTransistor}, Eqs.~\eqref{eq:LimitCyclePathway1}--\eqref{eq:DirectPathway} are (i) unfolding in a time dependent reference frame, (ii) containing terms deviating from classical TCMT \cite{Fan2003} and its nonlinear extensions \cite{Pedergnana2023,Pedergnana2024,Pedergnana2025}, and (iii) breaking Lorentz reciprocity \cite{Zhao2019}. Specifically, the oscillator equations \eqref{eq:LimitCyclePathway1}--\eqref{eq:LimitCyclePathway2} originate from an asymptotic averaging of a second order Liénard-type model for the pressure \cite{LimitCycleTransistor} and, therefore, (i) represent a slow time-scale evolution in a coordinate system moving in-phase with the incident wave, i.e., $\left\vert s^{\mathrm{+}}\right\rangle=\left\vert1\{2\}\right\rangle s\,,$ $s=\mathrm{const}$. Consequently, (ii) asymmetric detuning terms, $j(1-\Omega^2)/2\Omega$, and multiplicative forcing factors, $j/2\Omega$, follow. Finally, (iii) the coupling terms between incident wave and self-oscillator are assumed to depend on the direction of the energy flow [$\kappa\neq\tilde\kappa$] due to nonlinear effects.

The aforementioned considerations are required to adequately describe the physical phenomenon we observe, however, some further simplifications could be attempted. Specifically, the individual coefficients of the two oscillators [$f_{0,i}$, $\mu_i$, $\kappa_i$, $\tilde\kappa_i$ and $\varkappa$] could be, and in practice are, independent, however, imposing the simplifying assumptions $f_{0,1}=f_{0,2}=f_0$, $\mu_1=\mu_2=\mu$ and $|\varkappa|=|\kappa_1||\tilde\kappa_2|$ could still be useful. Indeed, those additional constraints allow us to isolate the essential features of the system with fewer parameters, while still remaining consistent with the experimental measurements.

In particular, Fig.~\ref{fig:ScatteringTheory} shows a comparison between experimental data [markers] and Eqs.~\eqref{eq:LimitCyclePathway1}--\eqref{eq:DirectPathway}, where the visualized trajectory through parameter space corresponds to an incident amplitude slightly varying with frequency due to reflections from the terminations [3D lines Fig.~\ref{fig:ScatteringExperiment}]. Solid and dotted lines depict solutions with independent [$\kappa_2=1.12\kappa$, $\tilde\kappa_2=1.12\tilde\kappa$] and identical [$\kappa_1=\kappa_2=\kappa$, $\tilde\kappa_1=\tilde\kappa_2=\tilde\kappa$] coupling coefficients, respectively, and the model parameters are reported in Tab.~\ref{tab:ReducedModelParameters}. Note that most details, including the jumps in $S_{12}$ and $S_{22}$ at $f=1184\,\mathrm{Hz}$ are well captured by allowing $\kappa_1\neq\kappa_2$, yet, even with the simplification $\kappa_1=\kappa_2=\kappa$ the main features of the synchronization driven process are quantitatively represented. For the sake of completeness the corresponding figure for the phases of the S-matrix elements are presented in the supplemental material \cite{supp}. Furthermore, the state of interest [Fig.~\ref{fig:ScatteringTheory}; blue shaded region] can be roughly estimated in simpler terms by the zeroth order expansion
\begin{equation}\label{eq:ApproximateS}
    \mathbf{S} \approx \mathbf{D} + \frac{\tilde\kappa}{s}\sqrt{\frac{\mu-2\mathrm{Re}(\varkappa)}{\mu+2\mathrm{Re}(\varkappa)}}
    \begin{bmatrix}
        1 & \frac{\mu+2\mathrm{Re}(\varkappa)}{\mu-2\mathrm{Re}(\varkappa)} \\
        0 & 2\frac{\mu+2\mathrm{Re}(\varkappa)}{\mu-2\mathrm{Re}(\varkappa)}
    \end{bmatrix}\,,
\end{equation}
where $\mathbf{D}\approx\mathbf{1}$, and the root corresponds to the unperturbed amplitude of the anti-phase mode, $|a_{1}|=|a_{2}|=|a_0^\mathrm{-}|$, or the inverse of the in-phase one, $|a_0^\mathrm{+}|=1/|a_0^\mathrm{-}|$. Crucially, $S_{21}\approx0$ due to superposition of anti-phase sources, while $S_{21}\approx|a_0^\mathrm{+}|\tilde\kappa/s$. Note that (i) for $s\rightarrow0$ and $\Omega=1$ the scattering matrix exhibits a singularity due to $|a_{1\{2\}}|\sim\mathcal{O}(1)$, (ii) for $\tilde\kappa\rightarrow0$ the direct process governs the behavior $\mathbf{S}=\mathbf{D}\approx\mathbf{1}$ and (iii) for $\varkappa\rightarrow0$ the phase difference between the whistles is undetermined and $S_{21}$ fluctuates without reaching a steady state.


\emph{Conclusion -- }
Above we have demonstrated that by utilizing the universal phenomenon of synchronization in weakly nonlinear limit cycle oscillators, we can break acoustic reciprocity in a manner that is both robust and tunable. Since the aforementioned mechanism depends only on the existence of coupled self-oscillators, it is generally independent of the specific physical implementation. Indeed, the same principles that govern the injection‐locking of microwave oscillators \cite{Perlman1970,Kurokawa1973} (e.g., a DC‐biased Gunn diode locking to an RF input) apply equally to our aeroacoustic whistles. In both cases a DC‐like bias establishes an autonomous oscillation, which can then be phase‐ and frequency‐locked by an external drive within the synchronization region (Arnold tongue) in parameter space \cite{Li2022,Navarro2022,Euler2024}. Moreover, synchronization based methods have been shown to transcend specific areas, e.g., by a direct parallel between a phonon \cite{Knunz2010} and microwave \cite{Otterstrom2020} laser. Therefore, in all domains supporting self-sustained oscillations, a synchronization-driven diode should be implementable by appropriately arranging the coupling and phase relations of its limit cycles. In conclusion, the presented synchronization-based paradigm may enable a new class of nonreciprocal devices across the electromagnetic, mechanical, and fluid-mechanical realms, with the potential of complementing classical resonance based methods with intrinsic loss-compensation, and straightforward reconfigurability through control of pumping and coupling parameters.

\bibliography{references}

\onecolumngrid
\appendix
\newpage

\begin{center} \large \bf 
	Supplemental Material\\
	for manuscript: Synchronization driven reciprocity breaking
\end{center}
by Alexander K. Stoychev,$^1$ Ulrich Kuhl,$^{1,2}$ and Nicolas Noiray$^1$\\
$^1$ CAPS Laboratory, Department of Mechanical and Process Engineering, ETH Zürich, 8092 Zürich, Switzerland\\
$^2$ Université Côte d’Azur, CNRS, Institut de Physique de Nice (INPHYNI), 06200, Nice, France\\

\renewcommand\thefigure{S\arabic{figure}} 
\setcounter{figure}{0}
\setcounter{page}{1}
\renewcommand{\thepage}{S\arabic{page}}
\setcounter{page}{1}
\setcounter{equation}{0}
\renewcommand\theequation{S\arabic{equation}}
\renewcommand{\appendixname}{Suppl.~material}
\renewcommand\thesection{\Alph{section}}
\vspace{1ex}
\hrule
\vspace{1ex}

\begin{SCfigure}[][h]
    \includegraphics[width=360pt]{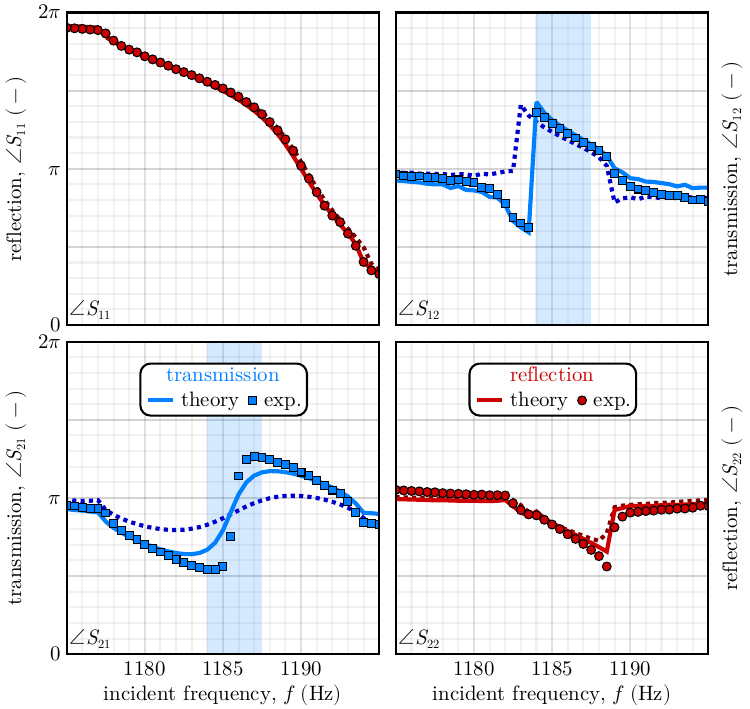}
    \caption{Phase of the scattering coefficients as a function of frequency, $f$, for an incident amplitude of $\left\vert\left\vert s_1^{\mathrm{+}}\right\rangle\right\vert\approx\left\vert\left\vert s_2^{\mathrm{+}}\right\rangle\right\vert\approx29\,\mathrm{Pa}$ and flow rates of $\dot{Q}_1=11\,\mathrm{l/min}$ and $\dot{Q}_2=11.25\,\mathrm{l/min}$, respectively. Left panels refer to incidence from port 1 [$S_{11}$ \& $S_{21}$], while right panels correspond to waves incident from port 2 [$S_{21}$ \& $S_{22}$; cf. Fig~\ref{fig:ExperimentalSetup}]. Blue colors represent transmission phase, with square markers depicting measurement points and solid (dotted) lines - theoretical calculations. Conversely, red colors denote reflection phase, with circular markers being measurements and solid (dotted) lines - solutions of Eqs.~\eqref{eq:LimitCyclePathway1}--\eqref{eq:DirectPathway}. Dotted lines correspond to $\kappa_2=\kappa_1=\kappa$ and solid ones to $\kappa_2=1.12\kappa$, where the latter is shown to be responsible for the phase flop and bandwidth widening in $S_{12}$. Note that the model represents the phase of the scattering coefficients faithfully up to a constant global phase. Specifically, the results here compensate for constant, frequency-independent phase-shift between theory and experiment by removing the average phase difference.}
    \label{fig:ScatteringTheoryPhase}
\end{SCfigure}

\twocolumngrid

\section{Scattering matrix phases}
Figure~\ref{fig:ScatteringTheoryPhase} shows comparison of the scattering matrix' phase between experimental data [markers] and Eqs.~\eqref{eq:LimitCyclePathway1}--\eqref{eq:DirectPathway}, where the visualized trajectory through parameter space corresponds to an incident amplitude varying slightly with frequency due to reflections from the terminations [cf.\ 3D lines in Fig.~\ref{fig:ScatteringExperiment}]. Solid and dotted lines depict solutions with independent [$\kappa_2=1.12\kappa$, $\tilde\kappa_2=1.12\tilde\kappa$] and identical [$\kappa_1=\kappa_2=\kappa$, $\tilde\kappa_1=\tilde\kappa_2=\kappa$] coupling coefficients, respectively, and the model parameters are the same as in Fig.~\ref{fig:ScatteringTheory}, i.e., the ones from Tab.~\ref{tab:ReducedModelParameters}. We can observe that most details, including the phase flip in $S_{12}$ at $f=1184\,\mathrm{Hz}$ are well captured by allowing $\kappa_1\neq\kappa_2$, with stronger deviations under the simplification $\kappa_1=\kappa_2=\kappa$. The good agreement of the phase is an additional evidence for the proposed modified temporal coupled mode theory. However, note that the model cannot necessarily account for global phase-shifts introduced by the nontrivial geometry of the setup, hence, the predictions are accurate up to a frequency-independent, global phase.

\end{document}